\documentclass[aps,prd,superscriptaddress,floatfix,showpacs,showkeys,preprintnumbers]{revtex4}
\usepackage[utf8]{inputenc}
\usepackage[T1]{fontenc}
\usepackage{slashed}
\usepackage{times}
\usepackage{amssymb,amsfonts,amsmath,amsthm}
\usepackage{dsfont,bbm}
\usepackage{dcolumn}
\usepackage{epsf}
\usepackage{graphicx}
\usepackage[caption=false]{subfig}
\usepackage{dsfont}
\usepackage{slashed}
\usepackage[active]{srcltx}
\usepackage[usenames]{color}

\begin{document}

\title{Novel effects in twist-three SSA}

\author{I.~V.~Anikin}
\email{anikin@theor.jinr.ru}
\affiliation{Bogoliubov Laboratory of Theoretical Physics, JINR,
             141980 Dubna, Russia}
\affiliation{Institut f\"ur Theoretische Physik, Universit\"at
Regensburg,D-93040 Regensburg, Germany}

\author{O.~V.~Teryaev}
\email{teryaev@theor.jinr.ru}
\affiliation{Bogoliubov Laboratory of Theoretical Physics, JINR,
            141980 Dubna, Russia}

\begin{abstract}
 We discuss the role of gluon poles and
the gauge invariance for the hadron tensors
of Drell-Yan and direct photon production processes with the transversely polarized hadron.
These hadron tensors are needed to construct the corresponding single spin asymmetries.
For the Drell-Yan process, we perform our analysis within both the Feynman and axial-type (contour)
gauges for gluons.
In both the Feynman and contour gauges, we demonstrate that the gauge invariance leads
to the need of the new (non-standard) diagrams.
Moreover, in the Feynman gauge, we argue the absence of gluon poles in the
correlators $\langle\bar\psi\gamma_\perp A^+\psi\rangle$
related traditionally to $dT(x,x)/dx$.
As a result, these terms disappear from the final
QED gauge invariant Drell-Yan hadron tensor.
For the direct photon production, by using the contour gauge for gluon fields,
we find that there are new twist-$3$ terms present in the hadron tensor of the considering process in
addition to the standard twist-$3$ terms.
\end{abstract}
\date{\today}
\maketitle

\section{Introduction}

We discuss the
direct photon production in two hadron collision where one hadron is  transversely polarized.
We present the hadron tensor for this process and study
the effects which lead to the soft breaking of factorization
(or the universality breaking) through
the QED and QCD gauge invariance \cite{AT-DPP}.
In a similar manner as in \cite{AT-10}, the special role is played by the contour gauge for gluon fields.

Moreover, we present the results of the
detailed analysis of hadron tensor in the Feynman gauge with the particular emphasis
on the QED gauge invariance \cite{AT-FG}.
Moreover, the results in the Feynman and contour gauges coincide if the
gluon poles in the correlators $\langle\bar\psi\gamma_\perp A^+\psi\rangle$ are absent.
This is in agreement with the relation between gluon poles and the Sivers function which
corresponds to the "leading twist" Dirac matrix $\gamma^+$. This also suggests that the
gluonic poles appear only in "physical" components of gluon field, coinciding with total ones in the axial gauge.


\section{Direct Photon Production}

\begin{figure}[t]
\centerline{
\includegraphics[width=0.3\textwidth]{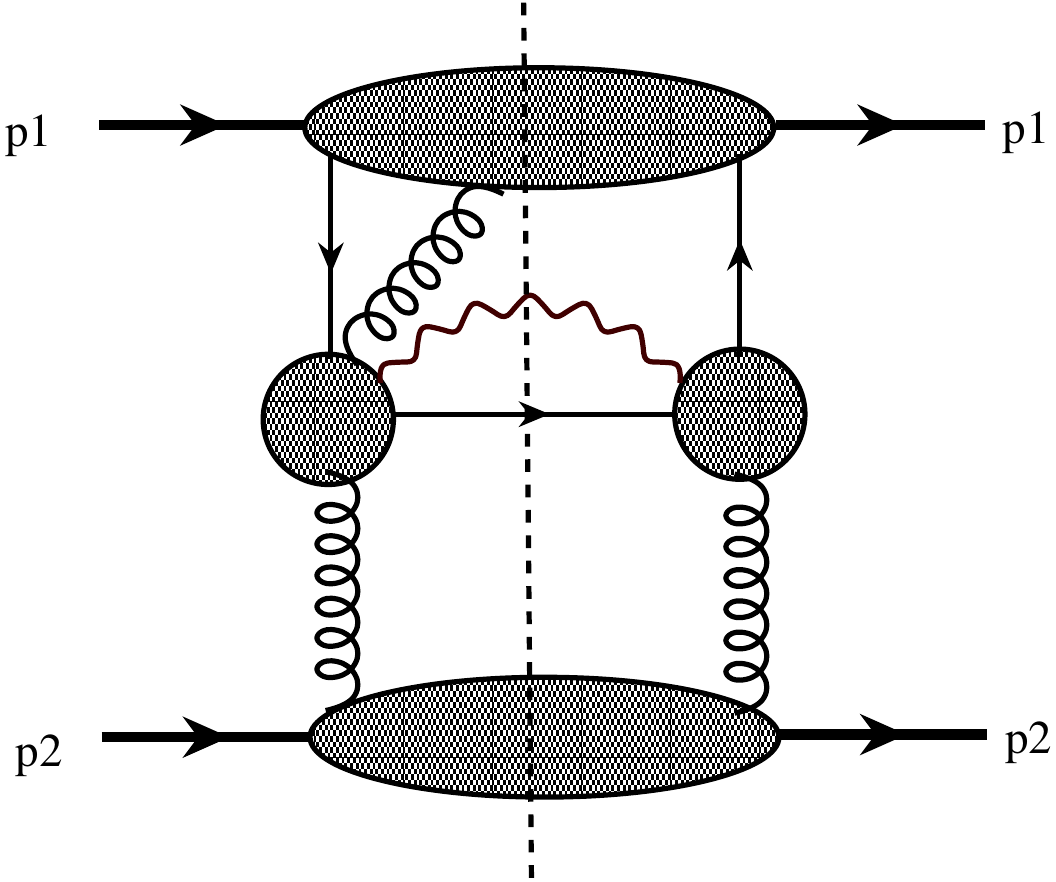}}
\caption{The Feynman diagram describing the hadron tensor of the
direct photon production.}
\label{Fig-DirPhot}
\end{figure}

We study the semi-inclusive process where the hadron with the transverse polarization
collides with the other unpolarized hadron to produce
the direct photon in the final state in:
\begin{eqnarray}
\label{process}
N^{(\uparrow\downarrow)}(p_1) + N(p_2) \to \gamma(q) + q(k) + X(P_X)\,.
\end{eqnarray}
For (\ref{process}) (also for the Drell-Yan process), the gluonic poles manifest.
We perform our calculations within a {\it collinear} factorization and, therefore,
it is convenient to fix the dominant light-cone directions as
\begin{eqnarray}
\label{kin-DY}
&&p_1 = \sqrt{\frac{S}{2}}\, n^*\, ,
\quad p_2 = \sqrt{\frac{S}{2}}\, n\,,\quad
\nonumber\\
&&
n^*_{\mu}=(1/\sqrt{2},\,{\bf 0}_T,\,1/\sqrt{2}), \quad n_{\mu}=(1/\sqrt{2},\,{\bf 0}_T,\,-1/\sqrt{2})\, .
\end{eqnarray}
Accordingly, the quark and gluon momenta $k_1$ and $\ell$ lie
along the plus dominant direction while the gluon momentum $k_2$ -- along the minus direction.
The final on-shell photon and quark(anti-quark) momenta can be presented as
\begin{eqnarray}
\label{Photon-Quark}
q = y_B\, \sqrt{\frac{S}{2}}\, n - \frac{q_\perp^2}{y_B \sqrt{2 S}}\, n^* + q_\perp\,,
\quad
k = x_B\, \sqrt{\frac{S}{2}}\, n^* - \frac{k_\perp^2}{x_B \sqrt{2 S}}\, n + k_\perp\,.
\end{eqnarray}
And, the Mandelstam variables for the process and subprocess are defined as
\begin{eqnarray}
\label{MandVar}
&&S=(p_1+p_2)^2, \quad T=(p_1-q)^2,\quad U=(q-p_2)^2,
\nonumber\\
&& \hat s =(x_1p_1 + yp_2)^2 = x_1 y S, \quad \hat t=(x_1p_1 -q)^2=x_1 T, \quad
\hat u =(q-yp_2)^2 = yU.
\end{eqnarray}
The amplitude (or the hadron tensor) of (\ref{process}) constructed by the contributions from
(i) the leading (LO) in particle number  diagrams: two diagrams with a radiation of the photon before (${\cal A}^{{\rm LO}}_1$)
and after (${\cal A}^{{\rm LO}}_2$)
the quark-gluon vertex with the gluon going to the lower blob, see the right side of Fig.~\ref{Fig-DirPhot};
(ii) the next-to-leading order (NLO) diagrams: eight diagrams constructed from
the LO diagrams by insertion of all possible radiations
of the additional gluon which together with the quark goes to the upper blob,
see the left side of Fig.~\ref{Fig-DirPhot}.
So, we have the hadron tensor related to the corresponding asymmetry:
\begin{eqnarray}
\label{SSA}
d\sigma^\uparrow - d\sigma^\downarrow \sim {\cal W}=
\sum\limits_{i=1}^{2}\sum\limits_{j=1}^{8}
{\cal A}^{{\rm LO}}_i \ast {\cal B}^{{\rm NLO}}_j\,.
\end{eqnarray}
Here, we will mainly discuss the hadron tensor rather than the asymmetry
itself. So, the hadron tensor as an interference between the LO and NLO diagrams,
${\cal A}^{{\rm LO}}_i \ast {\cal B}^{{\rm NLO}}_j$, can be presented by Fig.~\ref{Fig-DirPhot}
where the upper blob determines the matrix element of the twist-3 quark-gluon operator
while the lower blob -- the matrix element of the twist-2 gluon operator related to the
unpolarized gluon distribution.

As a result of factorization, we should reach the factorized form for the considered subject:
\begin{eqnarray}
\label{Fac-DY}
\text{Hadron tensor} = \{ \text{Hard part (pQCD)}\} \otimes
\{ \text{Soft part (npQCD)} \}\,.
\end{eqnarray}

We now dwell on the QCD gauge invariance of the hadron tensor
for the direct photon production (DPP).
First of all, let us remind that
having used the {\it contour gauge conception} \cite{AT-DPP, ContourG},
one can check that the representation
\begin{eqnarray}
\label{B-plus}
B^V_{+}(x_1,x_2)= \frac{T(x_1,x_2)}{x_1-x_2 + i\epsilon}\,
\end{eqnarray}
belongs to the gauge defined by $[x,\,-\infty]=1$, while
the representation
\begin{eqnarray}
\label{B-minus}
B^V_{-}(x_1,x_2)=\frac{T(x_1,x_2)}{x_1-x_2 - i\epsilon}\,
\end{eqnarray}
corresponds to the gauge that defined by $[+\infty,\, x]=1$.
In both (\ref{B-plus}) and (\ref{B-minus}), the function $T(x_1,x_2)$ related to the
following prametrization:
\begin{eqnarray}
\label{parT}
&&\langle p_1, S^T | \bar\psi(\lambda_1 \tilde n)\, \gamma^+ \,
\tilde n_\nu G^{\nu\alpha}_T(\lambda_2\tilde n) \,\psi(0)
|S^T, p_1 \rangle=
\nonumber\\
&&
\varepsilon^{\alpha + S^T -}\,(p_1p_2)\,
\int dx_1 dx_2 \, e^{i x_1\lambda_1+ i(x_2 - x_1)\lambda_2} \, T(x_1,x_2)\,.
\end{eqnarray}

Further, to check the gauge invariance,
we have to consider all typical diagrams, depicted in Fig.\ref{Fig-DirPhot}, that
correspond to the certain $\xi$-process (see, \cite{BogoShir}). Notice that, for the QCD gauge invariance, we have to assume
that all charged particles are on its mass-shells. That is, we will deal with only the physical gluons.

As a result of our calculation, the QCD gauge invariance (or the Ward identity) takes place provided only the
presence of the different complex prescriptions in gluonic poles dictated by the final or
initial state interactions:
\begin{eqnarray}
\label{QCD-illust}
  \begin{array}{rcl}
\hspace{-2cm}{\rm \bf FSI}&\Rightarrow& \frac{1}{\ell^+ + i\epsilon} \Rightarrow {\rm gauge}\,\,\,[+\infty^-,\, z^- ]=1
\Rightarrow \frac{T(x_1,x_2)}{x_1-x_2 - i\epsilon}\,\\
\\
\hspace{-2cm}{\rm \bf ISI}&\Rightarrow& \frac{1}{-\ell^+ + i\epsilon} \Rightarrow {\rm gauge}\,\,\,[z^-,\, -\infty^-]=1
\Rightarrow \frac{T(x_1,x_2)}{x_1-x_2 +i\epsilon}\,\\
  \end{array}\,\,\, \Bigg\}
\Rightarrow \textbf{QCD Gauge Invariance}\,,
\end{eqnarray}
where $\textbf{FSI}$ and $\textbf{ISI}$ imply the final and initial state interactions, respectively.

We now calculate the full expression for the hadron tensor
which involves both the standard and new contributions to the gluon pole terms.
After computing the corresponding traces and performing simple algebra within the frame we are choosing,
it turns out that the only nonzero contributions to the hadron tensor come from the following terms
\begin{eqnarray}
\label{diaH1}
d{\cal W}(\text{H1})&=&\frac{d^3 \vec{q}}{(2\pi)^3 2 E} \int \frac{d^3 \vec{k}}{(2\pi)^3 2 \varepsilon}
\delta^{(2)}(\vec{\bf k}_\perp+\vec{\bf q}_\perp)
\, \text{C}_2\, \int dx_1 dy \delta(x_1-x_B)\,\delta(y-y_B)\,  {\cal F}^g(y)\, \times
\nonumber\\
&&\int dx_2\, \frac{2 S^2\, x_1 \, y^2}{[x_2 y S + i\epsilon][x_1 y S + i\epsilon]^2}\,
\frac{\varepsilon^{q_\perp + S_\perp -}}{p_1^+}\, B^{V}_-(x_1,x_2)\,,
\end{eqnarray}
\begin{eqnarray}
\label{diaH7}
d{\cal W}(\text{H7})&=&\frac{d^3 \vec{q}}{(2\pi)^3 2 E} \int \frac{d^3 \vec{k}}{(2\pi)^3 2 \varepsilon}
\delta^{(2)}(\vec{\bf k}_\perp+\vec{\bf q}_\perp)
\, \text{C}_1\, \int dx_1 dy \delta(x_1-x_B)\,\delta(y-y_B)\, {\cal F}^g(y)\, \times
\nonumber\\
&&\int dx_2\, \frac{(-2) S\, T\, x_1\, (y-3y_B)}{[x_2 T + i\epsilon][x_1 T + i\epsilon]^2}\,
\frac{\varepsilon^{q_\perp + S_\perp -}}{p_1^+}\, B^{V}_+(x_1,x_2)\,,
\end{eqnarray}
\begin{eqnarray}
\label{diaD4}
d{\cal W}(\text{D4})&=&\frac{d^3 \vec{q}}{(2\pi)^3 2 E} \int \frac{d^3 \vec{k}}{(2\pi)^3 2 \varepsilon}
\delta^{(1)}(\vec{\bf k}_\perp+\vec{\bf q}_\perp)
\, \text{C}_1\, \int dx_1 dy \delta(x_1-x_B)\,\delta(y-y_B)\,\frac{2}{S}  {\cal F}^g(y)\, \times
\nonumber\\
&&\frac{2 S^2 \,x_1\, (y-2y_B)}{[x_1 T + i\epsilon]^2}\,
\frac{\varepsilon^{q_\perp + S_\perp -}}{2x_1p_1^+ + i\epsilon}\, \int dx_2\,B^{V}_+(x_1,x_2)\,,
\end{eqnarray}
and
\begin{eqnarray}
\label{diaH10}
d{\cal W}(\text{H10})&=&\frac{d^3 \vec{q}}{(2\pi)^3 2 E} \int \frac{d^3 \vec{k}}{(2\pi)^3 2 \varepsilon}
\delta^{(2)}(\vec{\bf k}_\perp+\vec{\bf q}_\perp)
\,\text{C}_3\, \int dx_1 dy \delta(x_1-x_B)\,\delta(y-y_B)\, {\cal F}^g(y)\, \times
\nonumber\\
&&\int dx_2\, \frac{2 T (x_1-x_2) (2 T+S y)}{[x_1 T + i\epsilon][x_2 T + i\epsilon][(x_1-x_2) y S + i\epsilon]}\,
\frac{\varepsilon^{q_\perp + S_\perp -}}{p_1^+}\, B^{V}_+(x_1,x_2)\,.
\end{eqnarray}
Here, $\text{C}_1=C_F^2 N_c$, $\text{C}_2=-C_F/2$, $\text{C}_3=C_F\, N_c \,C_A/2$.
The other diagram contributions disappear owing to the following reasons: (i) the $\gamma$-algebra gives
$(\gamma^-)^2=0$; (ii) the common pre-factor $T+yS$ goes to zero,
(iii) cancelation between different diagrams.

Analysing the results for the terms H1, H7, D4 and H10 (see Eqns. (\ref{diaH1})--(\ref{diaH10})),
we can see that
\begin{eqnarray}
\label{Fac2}
d{\cal W}(\text{H1}) + d{\cal W}(\text{H7}) + d{\cal W}(\text{D4}) =
d{\cal W}(\text{H10})\,.
\end{eqnarray}
In other words, as similar to the Drell-Yan process, the new (``non-standard") contributions
generated by the terms H1, H7 and D4 result again in the factor of $2$ compared to the
``standard" term H10 contribution to the corresponding hadron tensor.

\section{Drell-Yan process}

\begin{figure}[t]
\centerline{\includegraphics[width=0.3\textwidth]{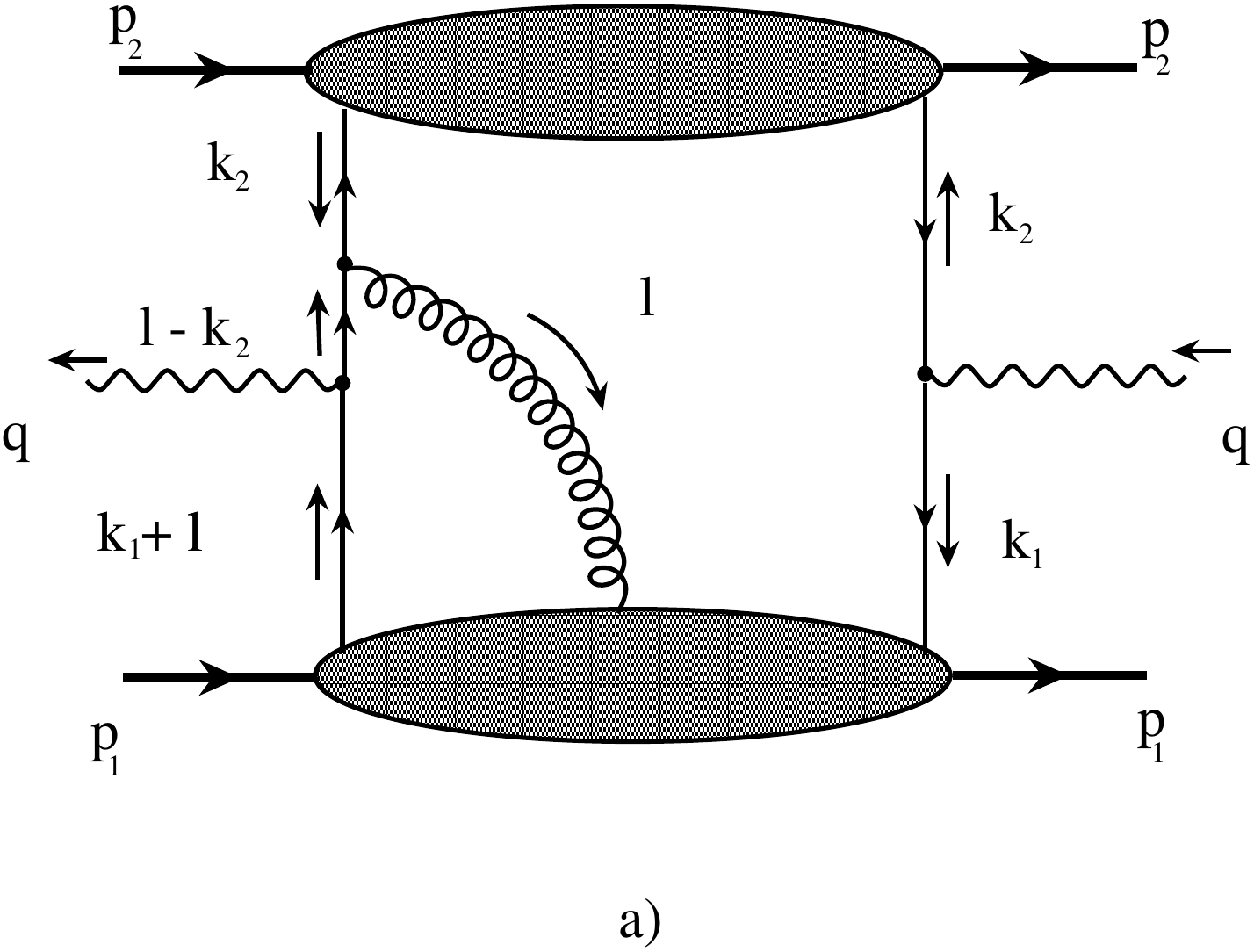}
\hspace{1.cm}\includegraphics[width=0.3\textwidth]{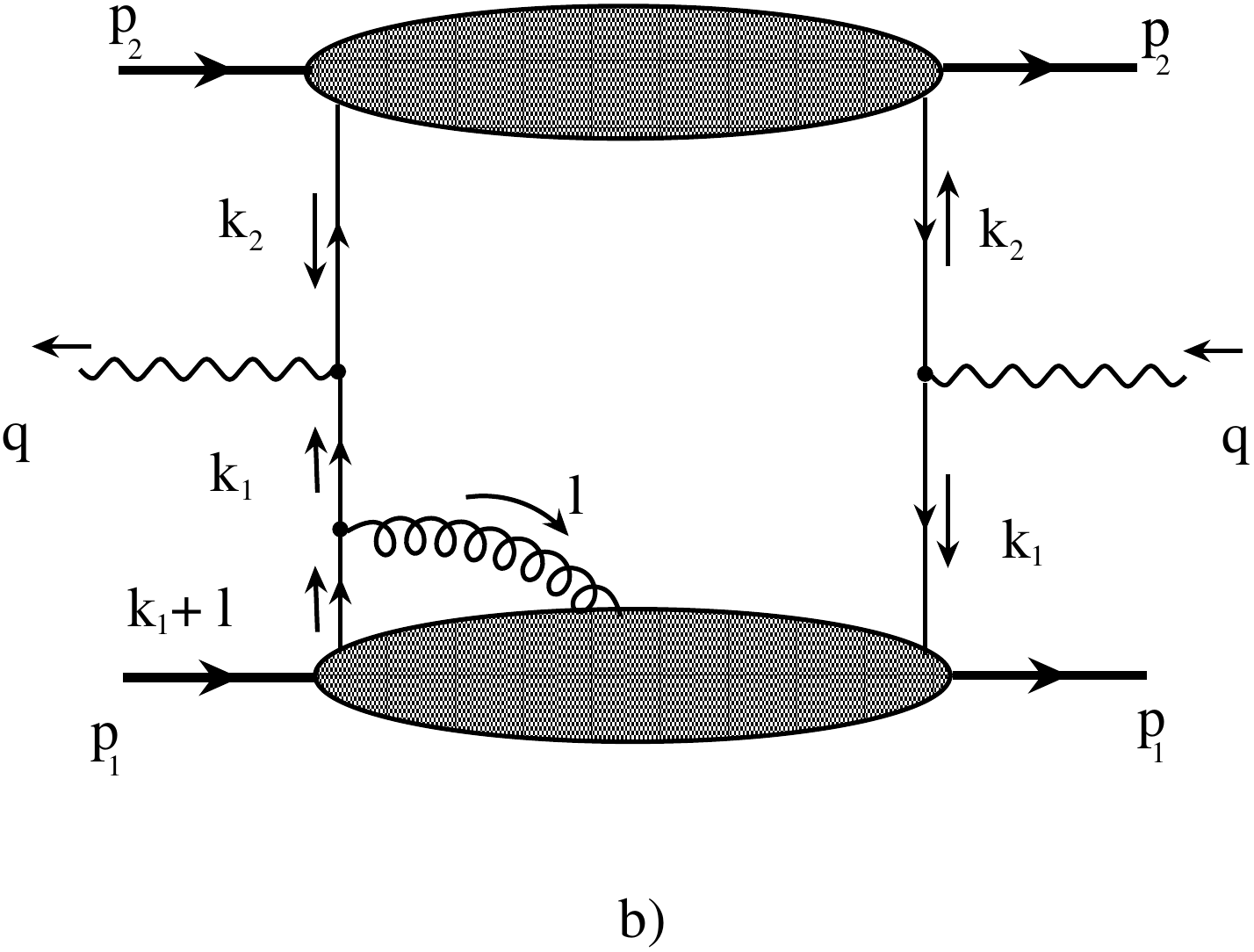}}
\caption{The Feynman diagrams which contribute to the polarized Drell-Yan hadron tensor.}
\label{Fig-DY}
\end{figure}

We now discuss the hadron tensor which contributes to the single spin
(left-right) asymmetry
measured in the Drell-Yan process with the transversely polarized nucleon (see Fig.~\ref{Fig-DY}):
\begin{eqnarray}
N^{(\uparrow\downarrow)}(p_1) + N(p_2) \to \gamma^*(q) + X(P_X)
\to\ell(l_1) + \bar\ell(l_2) + X(P_X).
\end{eqnarray}
Here, the virtual photon producing the lepton pair ($l_1+l_2=q$) has a large mass squared
($q^2=Q^2$)
while the transverse momenta are small and integrated out.
The left-right asymmetry means that the transverse momenta
of the leptons are correlated with the direction
$\textbf{S}\times \textbf{e}_z$ where $S_\mu$ implies the
transverse polarization vector of the nucleon while $\textbf{e}_z$ is a beam direction \cite{Barone}.

Since we perform our calculations within a {\it collinear} factorization,
it is convenient to fix  the dominant light-cone directions as
\begin{eqnarray}
\label{kin-DY}
&&p_1\approx \frac{Q}{x_B \sqrt{2}}\, n^*\, , \quad p_2\approx \frac{Q}{y_B \sqrt{2}}\, n,
\\
&&n^*_{\mu}=(1/\sqrt{2},\,{\bf 0}_T,\,1/\sqrt{2}), \quad n_{\mu}=(1/\sqrt{2},\,{\bf 0}_T,\,-1/\sqrt{2}).
\nonumber
\end{eqnarray}
So, the hadron momenta $p_1$ and $p_2$ have the plus and minus dominant light-cone
components, respectively. Accordingly, the quark and gluon momenta $k_1$ and $\ell$ lie
along the plus direction while the antiquark momentum $k_2$ -- along the minus direction.
The photon momentum reads (see Fig.~\ref{Fig-DY})
\begin{eqnarray}
q= l_1+l_2=k_1 + k_2= x_1 p_1 + y p_2\,+ q_T.
\end{eqnarray}

We work within the Feynman gauge for gluons.
After calculation of all relevant traces in the factorized hadron tensor and after some algebra, we arrive at the following
contributions for the unintegrated hadron tensor (which involves all relevant contributions except the mirror ones):
the standard diagram depicted in Fig.~\ref{Fig-DY}(a) gives us
\begin{eqnarray}
\label{DY-St}
&&\overline{\cal W}^{(\text{Stand.})}_{\mu\nu}
 + \overline{\cal W}^{(\text{Stand.},\,\partial_\perp)}_{\mu\nu}=\bar q(y)\,
\Bigg\{
- \frac{p_{1\,\mu}}{y}\,
\varepsilon_{\nu S^T -p_2}\, \int dx_2 \frac{x_1-x_2}{x_1-x_2+i\epsilon} B^{(1)}(x_1,x_2)
\nonumber\\
&&  -
\Big[ \frac{p_{2\,\nu}}{x_1} \varepsilon_{\mu S^T - p_2} + \frac{p_{2\,\mu}}{x_1} \varepsilon_{\nu S^T - p_2} \Big]
x_1\int dx_2 \frac{B^{(2)}(x_1,x_2)}{x_1-x_2+i\epsilon}
+ \frac{p_{1\,\mu}}{y} \,
\varepsilon_{\nu S^T - p_2}\, \int dx_2 \frac{B^{(\perp)}(x_1,x_2)}{x_1-x_2+i\epsilon}
\Bigg\}\,,
\nonumber
\end{eqnarray}
while the non-standard diagram presented in Fig.~\ref{Fig-DY}(b)
contributes as
\begin{eqnarray}
\label{DY-NonSt}
&&\overline{\cal W}^{(\text{Non-stand.})}_{\mu\nu}=
\bar q(y)
\frac{p_{2\,\mu}}{x_1}
\varepsilon_{\nu S^T -p_2}
\int dx_2 \Big\{ B^{(1)}(x_1,x_2) +
B^{(2)}(x_1,x_2)
\Big\}.
\end{eqnarray}
In Eqns.~(\ref{DY-St}) and (\ref{DY-NonSt}), all indices have to be treated as contravariant ones independently of
the real position in the formulae.
We also introduce the shorthand notation:
$\varepsilon_{A B C D}= \varepsilon_{\mu_1 \mu_2 \mu_3 \mu_4} A_{\mu_1} B_{\mu_2} C_{\mu_3} D_{\mu_4}$.
Moreover, the parametrizing functions are associated with the following correlators:
\begin{eqnarray}
\label{ParFunB1}
&&B^{(1)}(x_1,x_2)=\frac{T(x_1,x_2)}{x_1-x_2+i\varepsilon} \Longleftarrow
{\cal F}_2\Big[\langle p_1, S^T| \bar\psi(\eta_1)\, \gamma^+ \, A^T(z)\, \psi(0) | S^T,p_1 \rangle \Big]\,,
\\
\label{ParFunB2}
&& B^{(2)}(x_1,x_2) \Longleftarrow
{\cal F}_2\Big[\langle p_1, S^T| \bar\psi(\eta_1)\, \gamma^\perp \, A^+(z)\, \psi(0) | S^T,p_1 \rangle \Big]\,,
\\
\label{ParFunBperp}
&& B^{(\perp)}(x_1,x_2) \Longleftarrow
{\cal F}_2\Big[\langle p_1, S^T| \bar\psi(\eta_1)\, \gamma^+\, \big(\partial^\perp\, \, A^+(z)\big)\, \psi(0) | S^T,p_1 \rangle \Big]\,,
\end{eqnarray}
where ${\cal F}_2$ denotes the corresponding Fourier transformation.
Summing up all contributions from the standard and non-standard diagrams, we finally obtain
the expression for the unitegrated hadron tensor. We have
\nopagebreak
\begin{eqnarray}
\label{DY-ht-1}
&&\overline{\cal W}_{\mu\nu}=
\overline{\cal W}^{(\text{Stand.})}_{\mu\nu} + \overline{\cal W}^{(\text{Stand.},\,\partial_\perp)}_{\mu\nu}+
\overline{\cal W}^{(\text{Non-stand.})}_{\mu\nu}=
\bar q(y)\,
\Bigg\{ \Big[ \frac{p_{2\,\mu}}{x_1} - \frac{p_{1\,\mu}}{y} \Big] \,
\varepsilon_{\nu S^T -p_2}\, \int dx_2 B^{(1)}(x_1,x_2) +
\nonumber\\
&& \frac{p_{2\,\mu}}{x_1} \,
\varepsilon_{\nu S^T - p_2}\, \int dx_2 B^{(2)}(x_1,x_2) -
\Big[ \frac{p_{2\,\nu}}{x_1} \varepsilon_{\mu S^T - p_2} + \frac{p_{2\,\mu}}{x_1} \varepsilon_{\nu S^T - p_2} \Big]
x_1\int dx_2 \frac{B^{(2)}(x_1,x_2)}{x_1-x_2+i\epsilon} +
\nonumber\\
&& \frac{p_{1\,\mu}}{y} \,
\varepsilon_{\nu S^T - p_2}\, \int dx_2 \frac{B^{(\perp)}(x_1,x_2)}{x_1-x_2+i\epsilon}
\Bigg\}\,,
\end{eqnarray}
Notice that the first term in Eqn.~(\ref{DY-ht-1}) coincides with the hadron tensor calculated within the light-cone gauge
$A^+=0$.

Let us now discuss the QED gauge invariance of the hadron tensor.
From Eqn.~(\ref{DY-ht-1}),
we can see that the QED gauge invariant combination is
\begin{eqnarray}
\label{GI-comb}
&&{\cal T}_{\mu\nu}=\Big[ p_{2\,\mu}/x_1 - p_{1\,\mu}/y \Big] \,
\varepsilon_{\nu S^T -p_2},\,
\quad
q_\mu {\cal T}_{\mu\nu} = q_\nu {\cal T}_{\mu\nu}=0.
\end{eqnarray}
At the same time, there is a single term with $p_{2\,\nu}$ which does not have
a counterpart to construct the gauge-invariant combination
$p_{2\,\mu}/x_1 - p_{1\,\mu}/y.$
Therefore, the second term in  Eqn.~(\ref{DY-St}) should be equal to zero.
This also leads to nullification of the second term in Eqn.~(\ref{DY-NonSt}).
Now, to get the QED gauge invariant combination (see (\ref{GI-comb})) one has no the other way rather than
to combine the first terms in Eqns.~(\ref{DY-St}) and (\ref{DY-NonSt}) which justifies the treatment of gluon pole in $B^{(1)}(x_1,x_2)$
using the complex prescription.
In addition, we have to conclude that the only remaining third term in (\ref{DY-St}) should not contribute to SSA.

Moreover, we can conclude that, in the case with the substantial transverse component of the momentum,
there are no any sources for the gluon pole at $x_1=x_2$. As a result, the function  $B^{(2)}(x_1,x_2)$ has no gluon poles and,
due to T-invariance, $B^{(2)}(x_1,x_2)= - B^{(2)}(x_2,x_1)$, obeys $B^{(2)}(x,x) = 0$.
On the other hand, if we have $\gamma^+$ in the correlator, see (\ref{ParFunB1}), the transverse components of gluon momentum
are not substantial and can be neglected. That ensures the existence of the gluon poles for the function $B^{(1)}(x_1,x_2)$.
This corresponds to the fact that the Sivers function, being related to gluon poles, contains the "leading twist" projector $\gamma^+$.
So, we may conclude that in the Feynman gauge the structure $\gamma^+ (\partial^\perp A^+)$ does not produce the imaginary part and SSA as well.

Working within the Feynman gauge, we derive the
QED gauge invariant (unitegrated) hadron tensor for the polarized DY process:
\begin{eqnarray}
\label{DY-ht-GI}
\overline{\cal W}^{\text{GI}}_{\mu\nu}= \bar q(y)
\Big[ p_{2\,\mu}/x_1 - p_{1\,\mu}/y \Big]
\varepsilon_{\nu S^T -p_2} \hspace{-1.5mm}\int dx_2 B^{(1)}(x_1,x_2).
\end{eqnarray}
Further, after calculation the imaginary part (addition of the mirror contributions),
and after integration over $x_1$ and $y$,
the QED gauge invariant hadron tensor takes the form
\begin{eqnarray}
\label{DY-ht-GI-2}
W^{\text{GI}}_{\mu\nu}= \bar q(y_B)\,
\Big[ p_{2\,\mu}/x_B - p_{1\,\mu}/y_B \Big] \,
\varepsilon_{\nu S^T -p_2}\,  T(x_B ,x_B)\,.
\end{eqnarray}
This expression fully coincides with the hadron tensor which
has been derived within the light-cone gauge for gluons.
Moreover, the factor of $2$ in the hadron tensor,
we found within the axial-type gauge \cite{AT-10}, is still present in the frame
of the Feynman gauge.

\section{Conclusions and discussion}

the different prescriptions in the gluonic poles.
Moreover, the different prescriptions are needed to ensure the QCD gauge invariance.
This situation can be treated as a soft breaking of the
universality condition resulting in factorization breaking.
We find that the ``non-standard" new terms,
which exist in the case of the complex twist-$3$ $B^V$-function with the corresponding prescriptions,
do contribute to the hadron tensor exactly as the ``standard" term known previously.
This is  another important result of our work.
We also observe that this is exactly similar to the case of Drell-Yan process studied in \cite{AT-10}.

We argue the absence of gluon poles
in the correlators $\langle\bar\psi\gamma_\perp A^+\psi\rangle$
based on the light-cone dynamics.
At the same time, we also show that
the Lorentz and QED gauge invariance of the hadron tensor calculated within the Feynman gauge
requires the function $B^{(2)}(x_1,x_2)$ to be without any gluon poles.
This property seems to be natural from the point of view of gluon poles relation \cite{Boer:2003cm} to Sivers functions
as the latter is related to the projection $\gamma^+$. Note also that this may be understood as appearance of gluonic poles due to the physical components \cite{Hatta:2011ku} of gluonic fields.
Another support of this picture is provided by the fact, that imaginary phase in \ref{QCD-illust} allows to derive \cite{Teryaev:2015tza} Burkardt sum rules from conservation laws.

\section{Acknowledgements}

We would like to thank V.~Braun, I.O.~Cherednikov, A.V.~Efremov, D.~Ivanov and
L.~Szymanowski for useful discussions and correspondence.
This work is partly supported by the HL program.

\end{document}